\documentclass[aps,twocolumn,nopacs,floats,superscriptaddress]{revtex4}
\usepackage{graphicx}
\usepackage{dcolumn}
\usepackage{bm}
\usepackage{color}
\def\he4{$^4$He}
\def\hel3{$^3$He}
\def\Am3{\AA$^{-3}$}
\def\beq{\begin{equation}}
\def\eeq{\end{equation}}
\begin{document}


\author{Anatoly B. Kuklov}
\affiliation{Department of Physics \& Astronomy, College of Staten Island and the Graduate Center of
CUNY, Staten Island, NY 10314}

\author{Lode Pollet}
\affiliation{Arnold Sommerfeld Center for Theoretical Physics, University of Munich, Theresienstr. 37, 80333 M\"unchen, Germany}
\affiliation{Munich Center for Quantum Science and Technology (MCQST), Schellingstr. 4, 80799 M\"unchen, Germany}
\affiliation{Wilczek Quantum Center, School of Physics and Astronomy and T. D. Lee Institute, Shanghai Jiao Tong University, Shanghai 200240, China}

\author{Nikolay V. Prokof'ev}
\affiliation{Department of Physics, University of Massachusetts, Amherst, MA 01003, USA}

\author{Boris V. Svistunov}
\affiliation{Department of Physics, University of Massachusetts, Amherst, MA 01003, USA}
\affiliation{Wilczek Quantum Center, School of Physics and Astronomy and T. D. Lee Institute, Shanghai Jiao Tong University, Shanghai 200240, China}

\title{Supertransport by Superclimbing Dislocations in $^4$He: When All Dimensions Matter }
\begin{abstract}
The unique superflow-through-solid effect observed in solid \he4 and attributed to the quasi-one-dimensional
superfluidity along the dislocation cores exhibits two extraordinary features:
(i) an exponentially  strong suppression of the flow by a moderate increase in pressure, and
(ii) an unusual temperature dependence of the flow rate with no analogy to any known system
and in contradiction with the standard Luttinger liquid paradigm.
Based on {\it ab initio} and model simulations, we argue that the two features  are closely related:
Thermal fluctuations of the shape of a superclimbing  edge dislocation induce large,
correlated, and asymmetric stress fields acting on the superfluid core.
The critical flux is most sensitive to strong rare fluctuations and hereby acquires
a sharp temperature dependence observed in experiments.
\end{abstract}

\maketitle

{\it Introduction.} A pure (free of \hel3 impurities) but structurally imperfect crystal of \he4  is a highly non-trivial system demonstrating a variety of unique phenomena taking place at temperatures $T \lesssim 0.5$K that likely persist down to absolute zero:
(i) the superflow-through-solid (STS) 
\cite{Hallock,Hallock2012,Hallock2019,Beamish,Moses,Moses2019,Moses2020,Moses2021}, 
(ii) the anomalous isochoric compressibility (also known as the syringe effect), which is the thresholdless matter accumulation inside the solid in response to small chemical potential changes that always accompanies the STS \cite{Hallock}, and (iii) the giant plasticity \cite{Beamish_Balibar}. All three features are attributed to highly unusual and essentially quantum properties of dislocations.

The STS effect is explained by superfluidity in the cores of certain dislocations,
as established for both the screw and edge dislocations by {\it ab initio} path integral simulations in Refs.~\cite{screw,sclimb};
the original idea that dislocations in \he4 might have superfluid cores goes back to the work by Shevchenko \cite{shevchenko}.
The only existing scenario explaining the syringe effect is based on superclimb of edge dislocations \cite{sclimb}.
In contrast to the conventional climb assisted by pipe diffusion of thermally activated vacancies along the dislocation core~\cite{Lothe,Hull} (viable only at high temperature),  the superclimb is assisted by the superflow along the core.
The syringe effect persisting down to low temperatures when thermal activation is no longer possible, as well as
first-principle simulations of edge dislocations demonstrating superclimb, provide strong support to the minimalistic
unified scenario behind all phenomena based on the (quantum-)rough edge dislocations with superfluid cores.
\begin{figure}[!htb]
\includegraphics[width=0.9 \columnwidth]{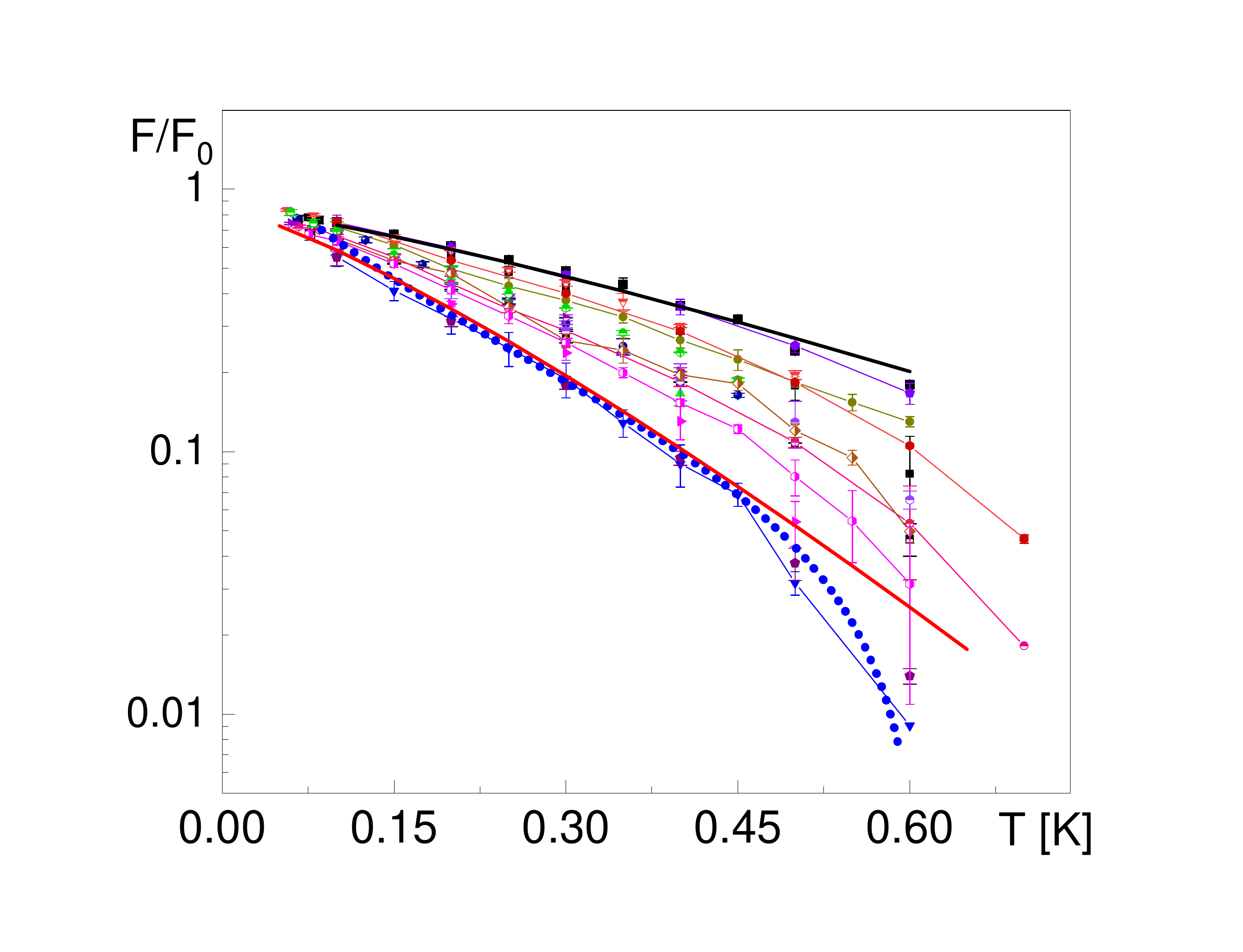}	
	\caption{Critical flux $F(T)$ [normalized by $F_0 = F(T\approx 0)$] for differently
grown \he4 samples. All data points connected by thin lines are a guide to the eye and are taken from
 Refs.~\cite{Moses,Moses2019}.
The dotted line represents the master curve, Ref.~\cite{Hallock}, fitting the data collected from multiple samples.
In contrast to Refs. \cite{Moses,Moses2019}, the data from Ref.~\cite{Hallock} shows no spread within $\sim$ 10\%.
All the data for $T<0.5$K are consistent with the stretched exponential law $\exp[-(T/T_{\alpha})^{\alpha}]$, $\alpha =$ 1--1.3,
predicted by our model: The thick solid lines are fits with $\alpha=5/4$ and $T_{5/4}\approx 0.20K$, $0.45K$ for the
lowest and the highest data sets, respectively, \cite{Moses,Moses2019}.}
 \label{A}
\end{figure}

Since dislocation cores are quasi-one-dimensional objects, it is natural to expect that their superfluid properties
fit the Luttinger liquid (LL) paradigm when at zero temperature $T$ the I-V curve is non-linear (sub Ohmic at small bias).
In agreement with the LL theory, {\it ab initio} simulations of dislocations with superfluid cores reveal
that LL parameters remain temperature-independent at $T \lesssim 0.5$K \cite{SM}.
These observations resulted in a widely shared point of view that supertransport of $^4$He atoms through the
dislocation network is described by a bosonic LL \cite{Hallock2019,Moses2021}.  
However, at low finite $T$, the initial part of the otherwise temperature-independent 
I-V curve in LL is supposed to acquire an Ohmic regime characterized by high conductivity diverging 
in the $T \to 0$ limit as a power law. In contrast,
experiments consistently 
observe a different mysterious temperature dependence
of the critical flux $F$ shown in Fig.~\ref{A} and, apparently, force one to look beyond
isolated dislocations and invoke properties of the dislocation network (cf. Ref.~\cite{trends}).

In this Letter, we argue that a single-dislocation scenario, where $F$ is simply a product of the critical current $J$ along one dislocation and a number of dislocations, is nevertheless possible by paying attention to the higher-dimensional nature of the problem---dislocations are defects of a three-dimensional crystalline order. 
In this context, two key ingredients become crucial: \\
(i)  thermal fluctuations of the dislocation shape, and \\
(ii) the exponential dependence of superfluid properties of the core
on moderate changes in pressure (more generally, the stress field around the dislocation core)---as observed in the experiment \cite{Moses}
and in our simulations (see below). \\
Their combination results in rare intermittent fluctuations with dislocation segments having strongly suppressed superfluid
stiffness in what otherwise is a robust superfluid along the core.
Phase slips in such regions limit the flux $F$. No other one-dimensional system is known to exhibit
a similar behavior.

Our scenario is supported by {\it ab initio} and model simulations and explains the leading behavior observed
in experiments, see Fig.~\ref{A}.
In particular, we explain finite size limitations of {\it ab initio} simulations and reveal strong suppression
of the superfluid stiffness with pressure and temperature. Of direct relevance to experiments is, however,
the critical current, which is out of reach for {\it ab initio} methods.
We corroborate the scenario by a simplified effective model, which captures, at least qualitatively,
the STS experiments that puzzled the community for over a decade. Given the spread of data for samples
with different geometry, size, and growth conditions, it is natural to expect that our modelling will
need further refinements: a group of samples from Ref.~\cite{Moses2019} follow
a stretched exponential law with parameters outside of the range supported by our current model.
In particular, our model does not capture the low-$T$ saturation behavior
seen in these samples. It also does not account for the sub-Ohmic dependence of the flux $F$ on the chemical potential
bias $\Delta \mu$ \cite{Hallock2012,Moses2019}, which we leave for future work.
The scope of the present Letter is hence limited to the unifying picture of the pressure and temperature
dependence over a wide temperature range.

\begin{figure}[!htb]
	\includegraphics[width=0.9\columnwidth]{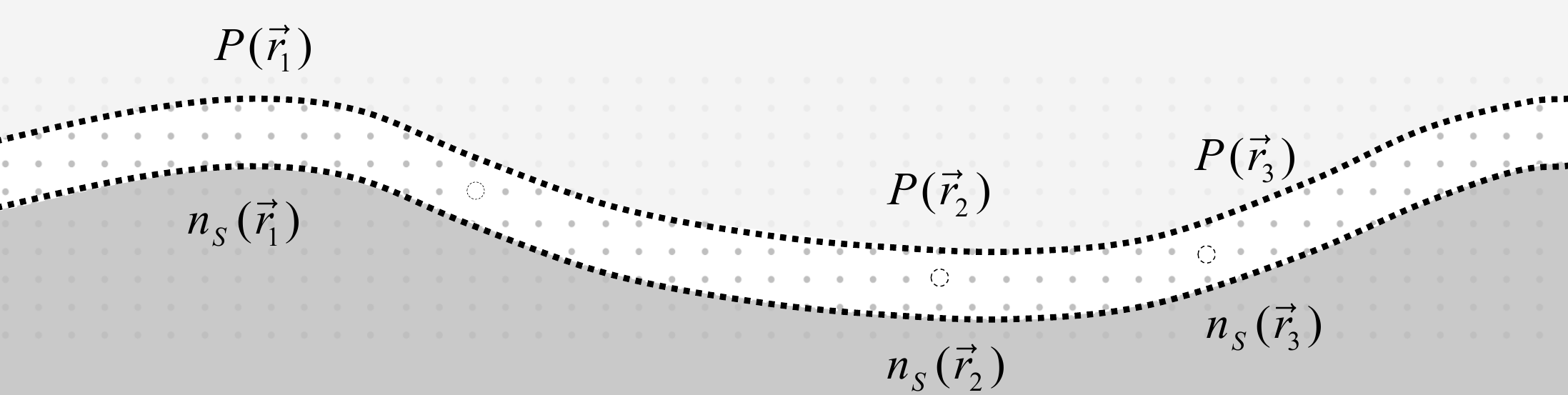}
\caption{Schematic visualization of shape fluctuations of the edge dislocation core in the climbing plane.
The incomplete atomic plane is shown by the dark area with the core at its boundary. 
The superfluid density is confined to the core (between the dashed lines) and its local value strongly depends
on the local pressure $P({\bf r})$ (or stress) .}
\label{cartoon}
\end{figure}

{\it Scenario.}
In solid $^{4}$He the motion of atoms/vacancies along the dislocation core is best described by
tunnelling in the periodic potential (see the sketch, Fig.~\ref{cartoon}) implying exponential
sensitivity of the superfluid properties to the potential strength.
%
Quantum roughness of the superfluid edge dislocation renders thermal fluctuations of their shape gapless
and, thus, anomalously large compared to the situation when the Peierls potential localizes the core within
a single potential minimum. Shape fluctuations of the edge produce inhomogeneous stress fields along the
core, which, in turn, modify the local superfluid response. Exponential sensitivity of tunneling
phenomena to external parameters, such as the local stress, amplifies the effect.

Imagine an instant shape of the dislocation line being quenched. Transport properties
of the resulting system are best described by the strong-disorder scenario \cite{WL00,WL0,WL}
when special attention is paid to the statistics of rare regions (outliers
in the pressure/stress distribution) creating ``bottlenecks" that might determine the current. 
The dynamical nature of thermal fluctuations does not allow us to take this analogy literally; e.g.,
phase transitions at finite $T$ are forbidden. Nevertheless, a sharp suppression of the superfluid density
$n_s$ and flux with temperature, see Fig.~\ref{A}, is possible.

The difference between the two properties is that  $n_s (T)$ dependence is a purely thermodynamic effect
based on a macroscopic number of connected regions with suppressed local superfluid response, or weak regions (WR),
while the critical current $J_c$ (with $F\propto J_c$) is not.
Because of the one-dimensional character of the system, $J_c$ is determined by a {\it single} WR
along the line allowing phase slips. The microscopic description of phase slips at WR goes beyond the scope of this work.
However, for qualitative comparison with the experiment all we need are the following two natural assumptions.
(i) $J_c$ is a product of the local superfluid density $n_s(x)$ and the local critical velocity at WR.
(ii) Phase slip is a ground-state, quantum-tunnelling phenomenon often called an instanton.
Smaller values of $n_s$ lead to a smaller instanton action and, correspondingly,
smaller critical velocities. Thus, we expect that $J_c$ scales as a certain power ($p > 1$)
of the local superfluid density $n_s$ at the WR.

 Here we do not consider thermal phase slips destroying the core superfluid in the thermally activated fashion \cite{Langer} at higher temperatures.
The corresponding activation energy $E_0$ is determined by $n_s(0)$ and the healing length $\xi$, and it can be estimated as
$E_0\approx \hbar^2 n_s(0)/(\xi m)$, where $m$ is the \he4 atomic mass. Our {\it ab initio} simulations give $E_0 \sim 5-10$K,
which is much higher than the typical energy scale $\sim 0.5$K observed experimentally in Refs. \cite{Hallock,Moses,Moses2019}.

\begin{figure}[!htb]
	\includegraphics[width=0.9 \columnwidth]{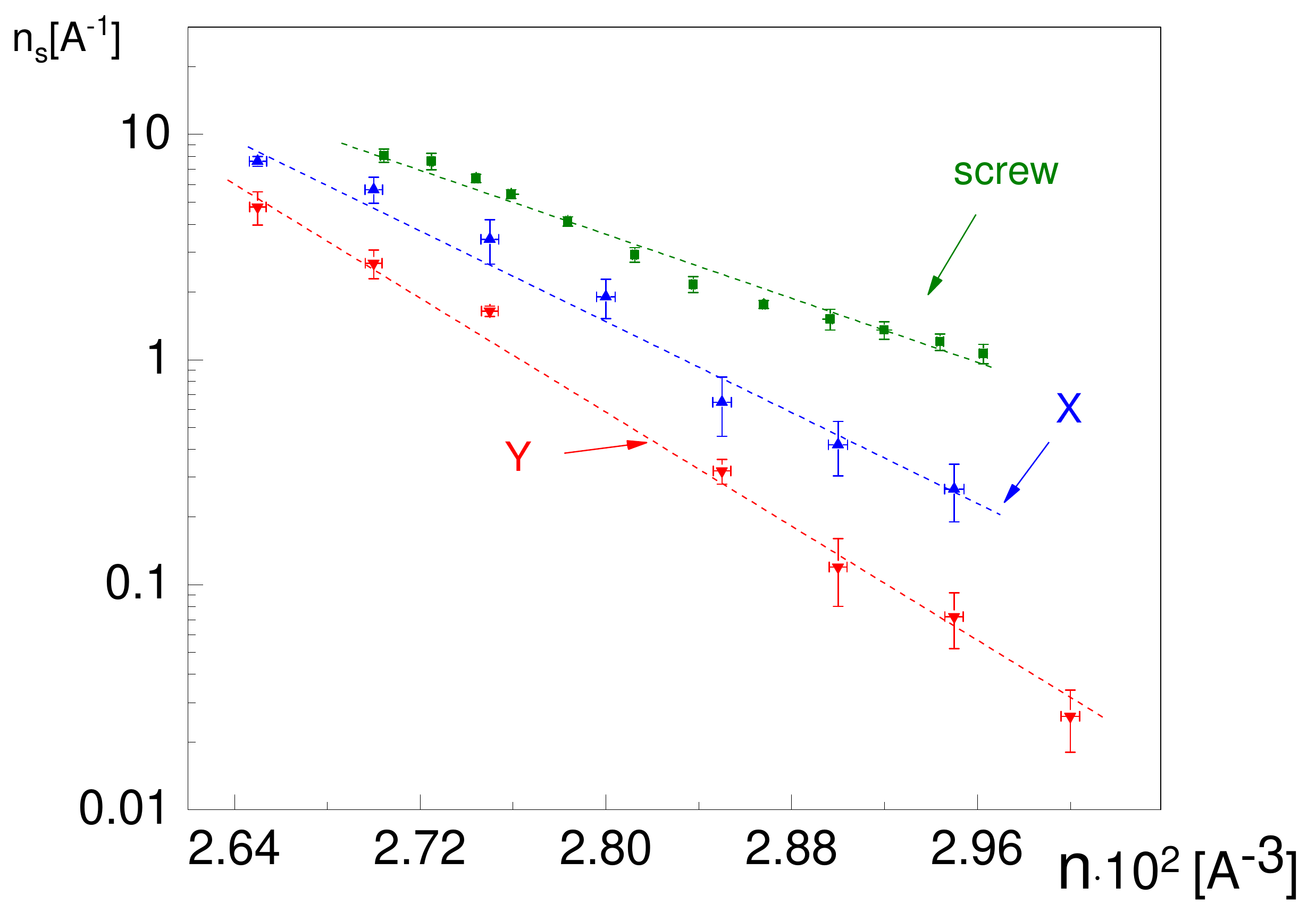}
	\caption{Exponential sensitivity of the superfluid density in the dislocation core, $n_s$,
to small changes in the 3D number density of the \he4 crystal, $n$. Three different dislocations were simulated at $T=0.25$K:
a screw dislocation and two edge dislocation partials along the [1,0,0] and [0,1,0] directions
labeled as X and Y, respectively. Each data point corresponds to a different sample at the corresponding density.
For more details see Ref.~\cite{SM}.}
\label{fig_ns}
\end{figure}

{\it Ab initio simulations.} The key assumption is the exponential sensitivity of $n_s$
to small changes in system parameters such as the crystal number density $n$. This was verified by {\it ab initio}
Worm Algorithm simulations \cite{WA} similar to those reported in Refs.~\cite{sclimb} but at higher densities
(see the Supplemental Material ~\cite{SM}): Fig.~\ref{fig_ns} clearly shows that $n_s$ can be suppressed
by nearly two orders of magnitude with only a $10 \% $ change in $n$.

The largest simulated system linear size $L$ cannot accommodate full-scale shape fluctuations
of the dislocation because $L$ does not satisfy the requirement
$L \gg D_0$, where $D_0 \sim 10 \AA$ is the core diameter. However,
rather than dealing with the shape fluctuations of a long dislocation line in an infinite ideal crystal,
we can study the statistics of position fluctuations of a short dislocation---of length $L \sim 20\AA ~\gtrsim D_0$---within
the box. Approximately, the dislocation can be viewed as composed of straight segments of length $L$,
moving with respect to each other and interacting by elastic forces.
The simulation box boundaries are formed by atoms with ``frozen'' spatial positions arranged to
enforce the topology of the dislocation inside the box (see Ref.~\cite{SM}).
As a result, the interior is under stress gradients with the characteristic scale $L$.
At the qualitative level, this arrangement mimics the effect of thermal shape fluctuations
and allows us to study how the dislocation segment explores the non-uniform stress landscape and
changes its superfluid properties depending on the position within the small box.

In the simulations, $n_s$ is calculated through the variance of the
winding number $W$ \cite{Pollock_Ceperley}:
$n_s =\hbar^{-2}mLT \, \langle W^2\rangle \, $,
\begin{figure}[!htb]
	\includegraphics[width=0.9 \columnwidth]{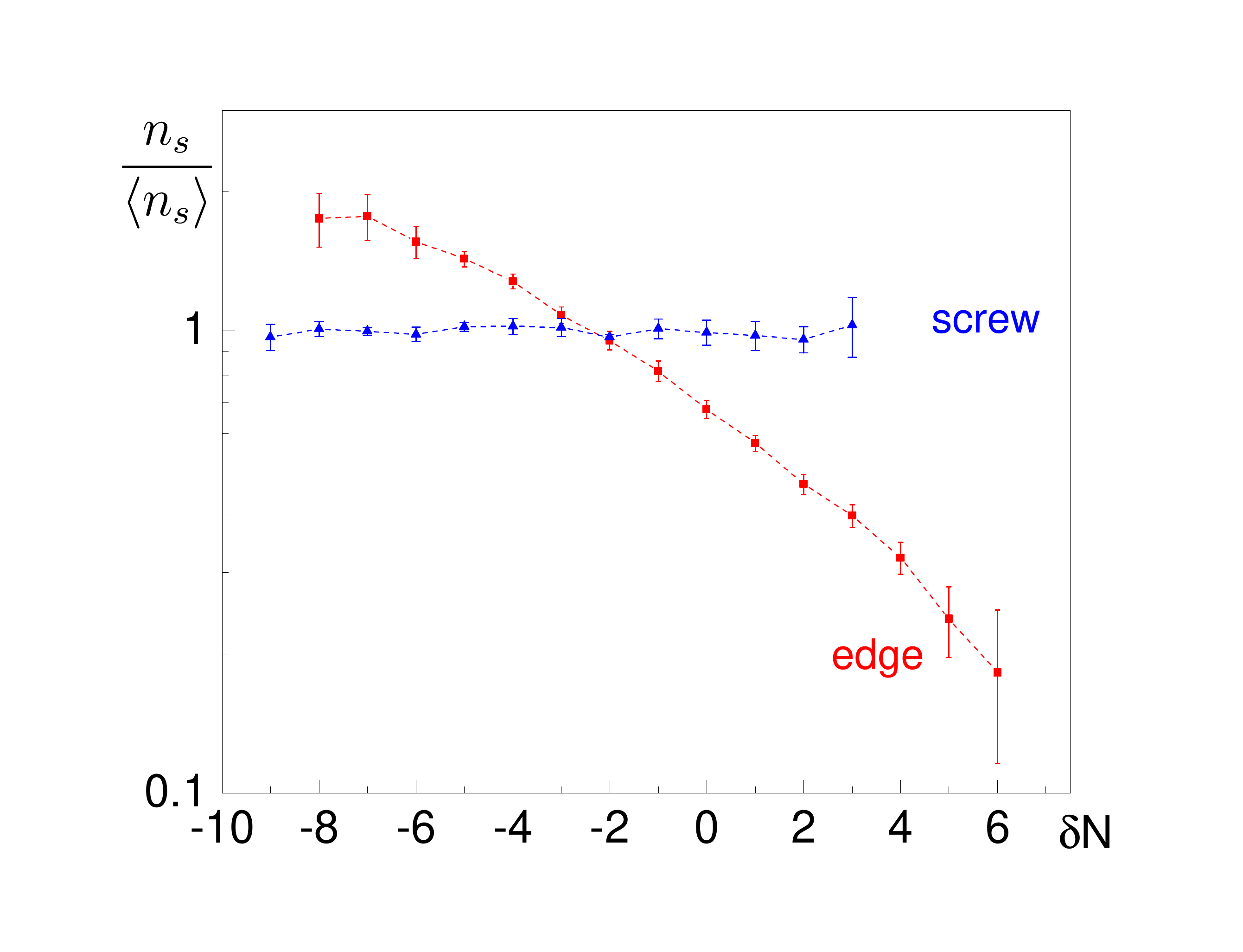}
	\caption{Superfluid density $n_s(N)$ as a function 
of the deviation $\delta N$ of the particle number $N$ from its equilibrium expectation value $\langle N \rangle$ in the simulation cell 
 for the superclimbing edge dislocation, at $T=0.5$K, and the screw dislocation with a superfluid core, at $T=1$K. The data are normalized by $\langle n_s \rangle$, which is $n_s$ averaged over $N$.}
\label{hist}
\end{figure}
where $m$ is the particle mass, and $\langle \ldots \rangle$ stands for averaging over an ensemble
of path-integral configurations. Large rare thermal fluctuations of the dislocation position are
statistically insignificant in $\langle W^2\rangle$ and this is why $n_s$ stays
$T$-independent in short samples \cite{SM}.
To reveal the effect of rare fluctuations one has to look at correlations between the $W^2$ and
dislocation core position within the 
simulation 
 cell, which can be readily done on the basis of the
one-to-one correspondence between the edge dislocation position and the particle number $N$.
This correspondence is the essence of the superclimb effect \cite{sclimb}:
the position of the core with respect to the crystal boundary determines the number of extra atoms belonging to the
incomplete atomic plane (the dark area in Fig.~\ref{cartoon}). Accordingly, fluctuations of the edge dislocation
imply fluctuations of $N$. Note that there is no such a relation for the screw dislocation. 
Since $N$ is a constant of motion, its fluctuations  are thermal, ensuring that we are studying
finite-$T$ rather than zero-point effects.

In the numerical protocol, the parameters are chosen so that $N$ experiences substantial fluctuations.
The statistics of $\langle W^2\rangle$ is then collected separately for each particle number $N$.
The corresponding quantity is denoted as $\langle W^2\rangle_N$ and defines the $N$-dependent superfluid density
$ n_s(N) = \hbar^{-2}mLT  \langle W^2\rangle_N$. If our scenario is correct, dramatic changes in
$n_s(N)$ between typical and rare values of $N$ should occur. The results are presented in Fig.~\ref{hist}.
For large deviations of $N$ from its expectation value $\langle N \rangle$ 
(that is, for large deviations of
the core from its equilibrium position within the simulation cell), 
we observe significant (by a factor of $\sim 5$)
changes in $n_s(N)$ (as compared with $n_s$ averaged over $N$) for a superclimbing edge dislocation.
In contrast, the screw dislocation with the superfluid core experiencing similar fluctuations in $N$
(which, however, cause no core displacement) 
demonstrates no dependence of $n_s(N)$ on $N$.  
This dramatic difference nicely illustrates the
key aspect of our scenario for edge dislocations---the dependence on $N$ is not due to particle density
fluctuations within the superfluid core (as is the case for the screw dislocation),
but due to the modification of the local crystalline environment around the climbing edge dislocation
in the presence of the stress field gradients.
The same conclusion follows from the direct comparison between the Figs.~\ref{hist} and \ref{fig_ns}
where both screw and edge dislocations demonstrate the exponential suppression with the crystal density $n$.

{\it Model simulations}. Qualitatively, the shape fluctuations of the superclimbing edge dislocation and their effect
on the superflow along its core can be studied within the simplified effective  model of an isotropic string
in two dimensions. That is, we ignore the presence of the crystalline lattice, which induces anisotropy and
the Peierls potential. The effective Hamiltonian reads (in the units when the interatomic distance $b$ and the ratio $\hbar/m$ equal to unity)
\beq
H[\varphi,y] = {\textstyle{1 \over 2}} \int_0^L dx \left\{n_s(y'') [V_0 + \varphi' ]^2 + G y'^2\right\} \, ,
\label{E}
\eeq
where the superfluid phase field, $\varphi(x)$, and the dislocation displacement field, $y(x)$,
are canonically conjugated variables. The term $\propto G$ is the elastic deformation energy.
We consider the limit of small deviations of the line from its equilibrium $y(x)=0$, that is, $|y'(x)|<1$.
Also, the line curvature $y''(x)$ must be much smaller than $1/D_0$.
The kinetic energy part contains the average flux velocity $V_0$, and the local superfluid density
\beq
n_s(y'') =n_0\exp(-gy''),
\label{exp}
\eeq
which depends exponentially on the shape fluctuations through the line curvature $y''$
(see Refs. \cite{Lothe,SM} for additional details). It is consistent with the dependence
on pressure/density observed in the {\it ab initio} simulations presented in Figs.~\ref{fig_ns}--\ref{hist}.

Quantum mechanical simulations of Eq.~(\ref{E}) at finite $V_0$ suffer from the sign problem.
However, for purposes of qualitative analysis it is sufficient to consider
the classical version of (\ref{E}) with explicit temperature-dependent ultraviolet cutoff, $\Delta x$,
on the wave lengths of excited modes for which quantization effects can be neglected.
We implement the cutoff by working with the discretized version of (\ref{E}) of linear size $\tilde{L} = L/\Delta x \gg 1 $
and the non-compact field $\varphi$. Given that the spectrum of excitations is parabolic
at small momenta \cite{sclimb}, $\omega = \sqrt{n_0 G} q^2$, \cite{sclimb}, we have
$\Delta x \approx (n_0G)^{1/4}/T^{1/2}$. This treatment is valid as long as $\Delta x \gg D_0$.

Brute-force classical simulations of (\ref{E}) suffer from severe slowing down because optimal WR configurations
introduce strong and highly non-local correlations between the fields $y$ and $\varphi$: A bump in the former is
accompanied by a large gradient in the latter at the bump location with reduced $\varphi$-gradients everywhere else.
Stochastic sampling of optimal WR configurations thus requires an enormous number of elementary local moves.
However, for purposes of the semi-quantitative analysis the slowing down problem can be solved by implementing
the least-energy approximation for the field $\varphi (x)$  when $\varphi (x)$ nothing but the optimal solution
for a given configuration of $y(x)$. In practice, this approximation reduces to an effective energy functional
that depends only on $y(x)$, which is then sampled stochastically.
The optimal phase field solution corresponds to the constant current condition:
\beq
J =   n_s(y''(x)) [V_0 + \varphi'(x) ] ,  \quad  dJ/dx=0 \, .
\label{J}
\eeq

\begin{figure}[!htb]
	\includegraphics[width= 1.0\columnwidth]{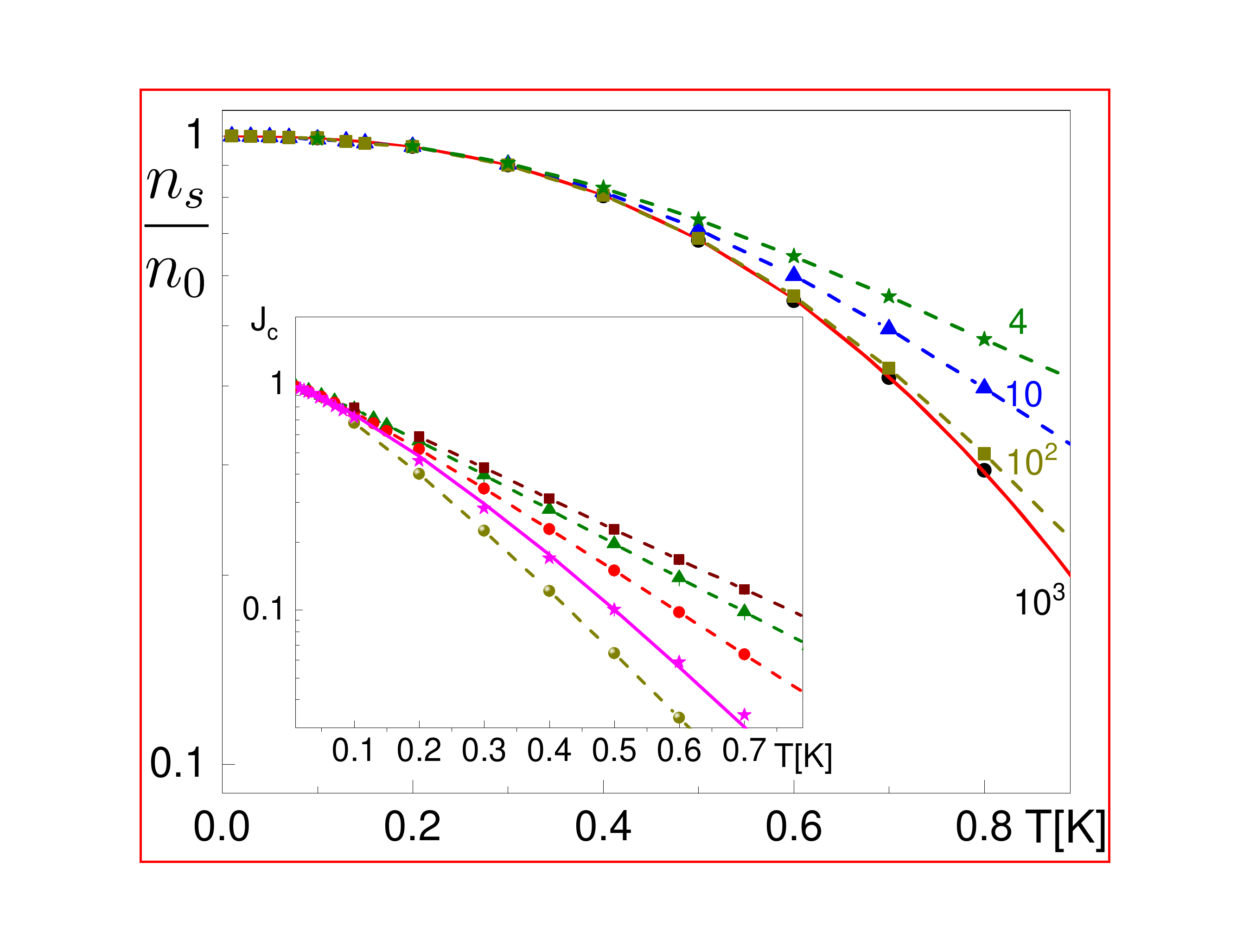}
\vskip -0.5cm	
\caption{ Superfluid density $n_s$ (dashed lines are guides to an eye)
for linear sizes $\tilde{L}$ specified next to each data set.
The solid line is the fit by $\exp[-(T/T_\alpha)^{\alpha}]$ with $\alpha=5/2$ and  $T_{\alpha}=0.737 \pm 0.005$.
Inset: The  critical current $J_c$ (normalized to unity at $T=0$) for $\tilde{L}=4, 10, 10^2, 10^3, 10^4, 10^7$ increasing from the highest to the lowest data set.
The solid line is the fit of the $\tilde{L}=10^4$ data by $J_c \propto \exp[-(T/T_\alpha)^{5/4}]$ with $T_\alpha =0.257 \pm 0.01 $.
All data for $J_c$ are consistent with $\alpha \approx 5/4 \pm 0.1$ for $0<T<0.6$ and the spread is due to
$T_\alpha$ varying by a factor $\sim 2$ (cf. Fig.~\ref{A}).}
\label{fig_nre}
\end{figure}

The superfluid density $n_s$ is computed from $n_s=\langle J \rangle/V_0$ in the $V_0 \to 0$ limit.
The critical flux $F \propto J_c$ cannot be determined from equilibrium Monte Carlo simulations.
Nevertheless, the origin of its $T$-dependence can be traced back to the statistics of the WR
through the proposed dependence of $J_c$ on superfluid density in WR
\beq
J_c \propto \langle n_s^p \rangle_{\rm WR} \propto \langle \exp(-pg y'') \rangle_{\rm WR} \,.
\label{JC}
\eeq
Here $ \langle ... \rangle_{\rm WR}$ denotes averaging over the WR.
(Note that an alternative interpretation in terms of the Landau criterion for stability of the
superflow is also possible \cite{SM}).
The results of the simulations for $n_s$ and $J_c$ with $p=2$ are presented in Fig.~\ref{fig_nre}
(see also Fig.~\ref{A}),
with further technical details delegated to the Supplemental Material~\cite{SM}.
\vskip -0.5cm

{\it Conclusions and outlook.}
Superflow-through-solid experiments exhibit a highly unusual temperature dependence of the critical flux (see Fig.~\ref{A})
that does not conform with the standard Luttinger liquid theory expected to work within the otherwise consistent and numerically corroborated picture of superfluidity confined to dislocation cores. Our scenario provides an explanation: the one-dimensional superfluid channel is an inseparable part of the crystaline host and its shape fluctuations induce pressure/stress and density fluctuations modifying properties of the edge dislocation core (see Fig.~\ref{hist}). Since superfluidity comes from tunneling
motion of core atoms in the periodic crystal potential the net result is an exponential dependence of $n_s$ on small
changes in $P$, and, consequently, 
in combination with the thermal fluctuations---on $T$.

The long-wave physics of rough superfluid edge dislocations is captured by the effective model (\ref{E}). It
reveals the effect of rare shape fluctuations on the superfluid density and emphasizes a nearly classical nature
of these fluctuations up to two quantum effects: (i) the UV cutoff on the wavelength of shape fluctuations
and (ii) the value of the largest ground-state flux limited by phase slips (this quantum-tunneling phenomenon
still remains to be understood).

The sharp temperature dependence of the critical flux is attributed to a single weakest region along the dislocation.
Phenomenological treatment based on the assumption that the critical flux scales as a certain power (larger than one) of the local superfluid density at the weakest element allows us to reproduce experimental results at $T \leq 0.5K$, see (Fig.~\ref{A}).
At higher temperature the data in Fig.~\ref{A} decay faster; this can be accounted for by considering the contribution of bulk phonons
to shape fluctuations, which leads to $\ln (F/F_0) \sim -T^2$~\cite{SM}.

The quantum treatment
of (\ref{E}) is of fundamental interest on its own because it describes a new type of the quasi-1D superfluid---not accounted for by the Luttinger Liquid paradigm.

The inset in Fig.~\ref{fig_nre}  demonstrates weak (logarithmic) suppression of the critical flux  with increasing the dislocation length. A similar suppression of the flux with the sample size has been reported in Ref.~\cite{Moses2019}. This motivates further experimental studies of the STS effect in single crystals with variable distance between the Vycor ``electrodes." If crystals can be grown with predominantly screw dislocations, the genuine
Luttinger-liquid behavior can be revealed through the temperature independent supercritical flux (at low temperature). 

\begin{acknowledgments}
We thank Robert Hallock and  Moses Chan for useful discussions and sharing the experimental data.
This work was supported by the National Science Foundation under the grants
DMR-2032136 and DMR-2032077.
We acknowledge the support from the CUNY High Performance Computing Center.
LP received funding from the European Research Council (ERC) under the European Union's Horizon 2020 research
and innovation programme (agreement No 771891 QSIMCORR) and the Deutsche Forschungsgemeinschaft
(DFG, German Research Foundation) under Germany's Excellence Strategy -- EXC-2111 - 390814868.
\end{acknowledgments}

\vskip 1cm

{\Large \bf Supplemental Material}
\section{Dislocations with superfluid cores in solid \he4}\label{sec1}
Superfluidity of the edge dislocation with its Burgers vector $\bf b$ along the hcp axis has been found numerically in Ref.~\cite{sclimb}. This dislocation splits into two partials separated by the basal fault of the type $E$ (see in Refs.~\cite{Lothe}) 
and each partial is characterized by half of the Burgers vector $\bf b$. 
Another partial, which we call here as $E_1$, is not related to any fault.  Both partials, $E$ and $ E_1$, were simulated by the worm algorithm \cite{WA}. There are also two more partial dislocations with the same (1/2 of) Burgers vector at the 
edges of two other hcp faults, $I_1$ and $I_2$  (see Refs.~\cite{Lothe}), which will be analyzed elsewhere. 
\begin{figure}[!htb]
	\includegraphics[width=0.9 \columnwidth]{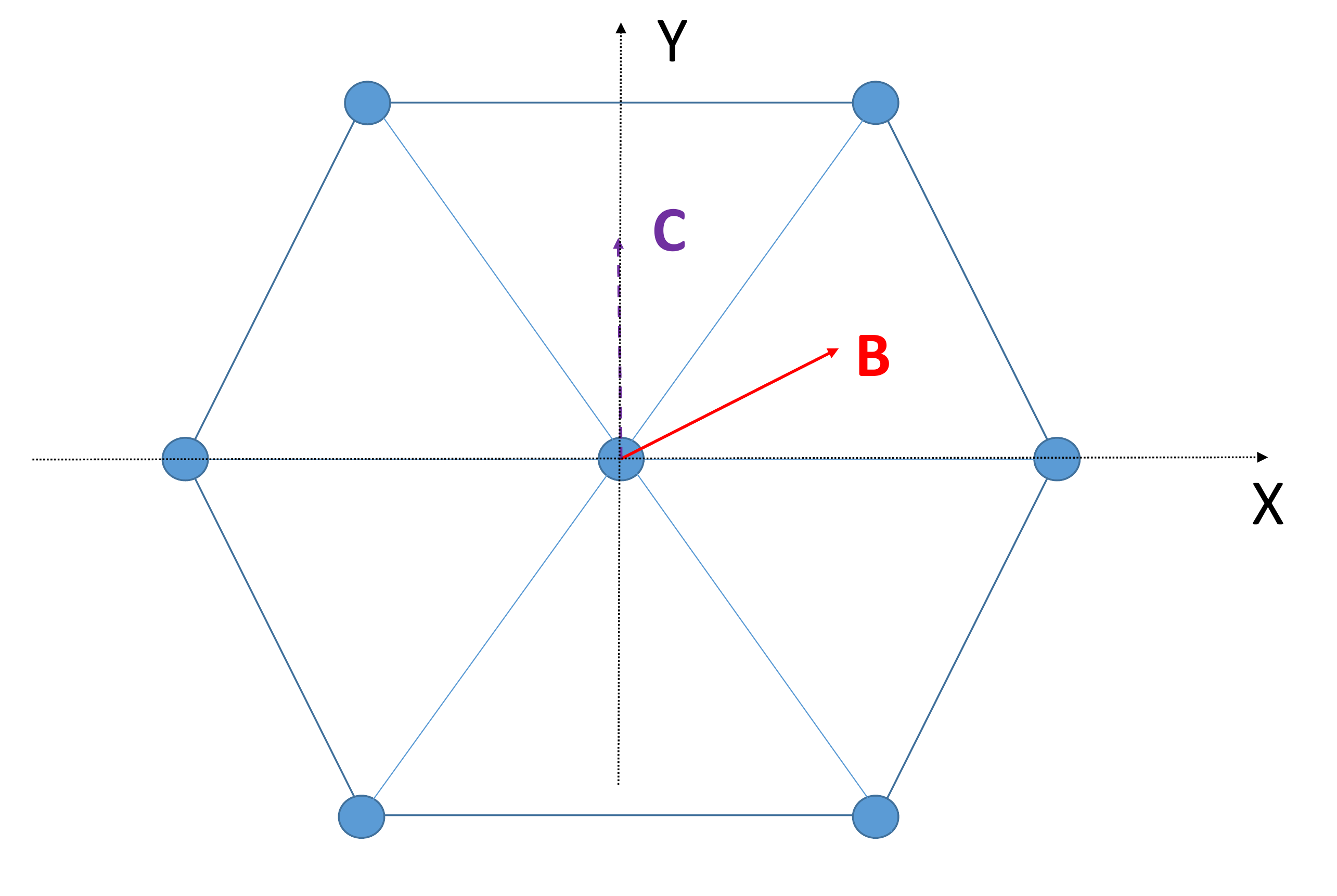}
	\caption{ (Color online) Triangular lattice of atoms in the A layer (located at, say, $z=0$). The adjacent B layer is shifted with respect to A by the vector ${\bf B}= (1/2, 1/\sqrt{12}, \sqrt{2/3}$) (solid arrow) in units of the closest distance between the atoms. The hcp structure is formed by the periodic sequence ABABAB.... The the fcc structure is formed by the sequence ABCABC... with the C layer being shifted by the vector ${\bf C}=(0, 1/\sqrt{3}, \sqrt{8/3})$ (dashed arrow).  }
\label{fig1}
\end{figure}
\begin{figure}[!htb]
\includegraphics[width=0.9 \columnwidth]{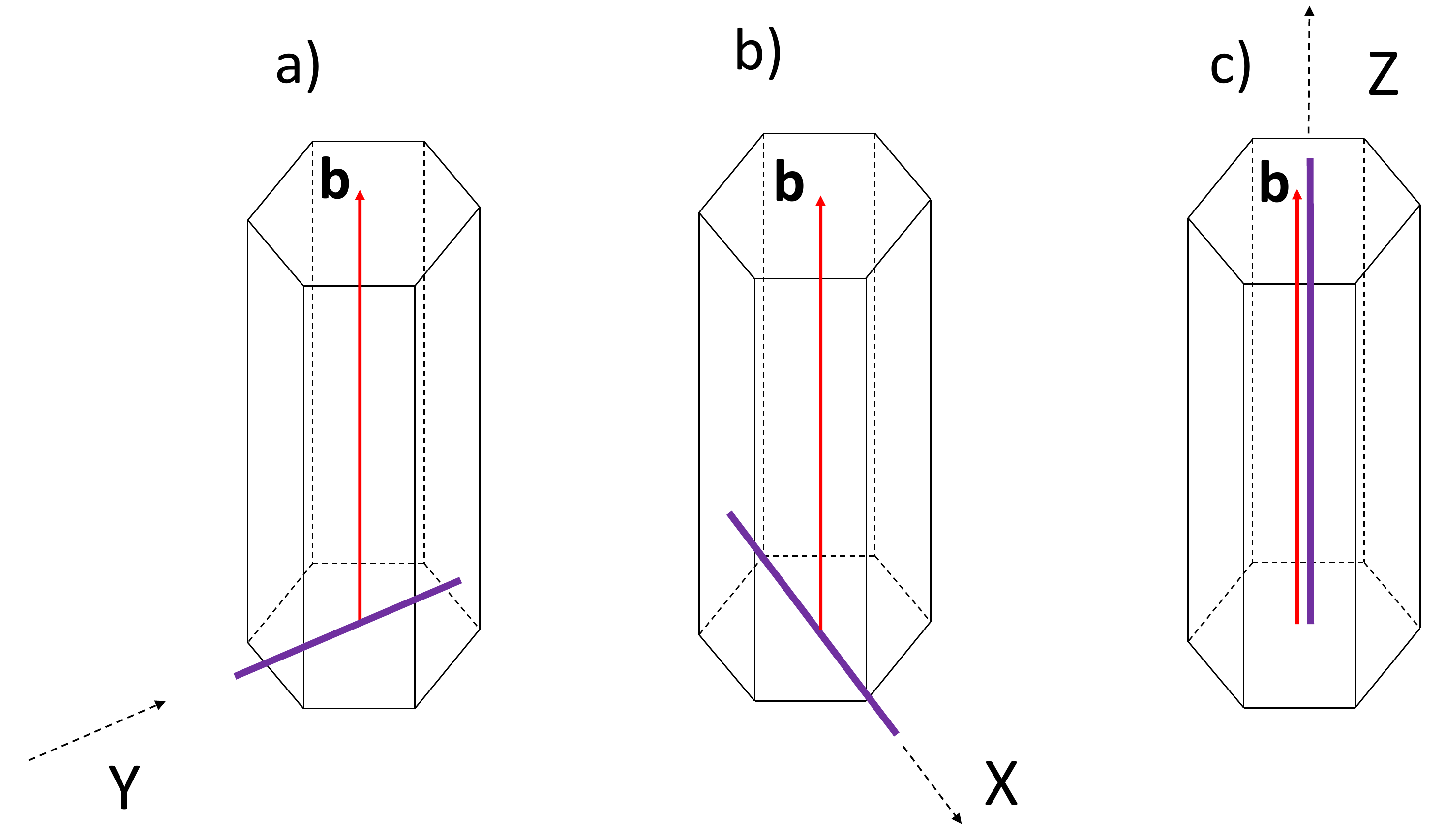}
\caption{ (Color online) Full edge dislocations  with the superfluid cores (thick solid lines) in the {\it hcp} solid \he4.
There are three orientations of the cores, which were analyzed in this work: 
(a) Edge dislocation with the core along the  [0,1,0] direction (Y-axis); 
(b) Edge dislocation with the core along the  [1,0,0] direction (X-axis); 
(c) Screw dislocation along the [0,0,1] direction. 
The solid (red) arrow along the hcp symmetry ($\hat{z}$)-axis shows the Burgers vector 
${\bf b}=(0,0,\sqrt{8/3})$ for all three dislocations.}
\label{fig2}
\end{figure}
\begin{figure}[!htb]
\includegraphics[width=0.9 \columnwidth]{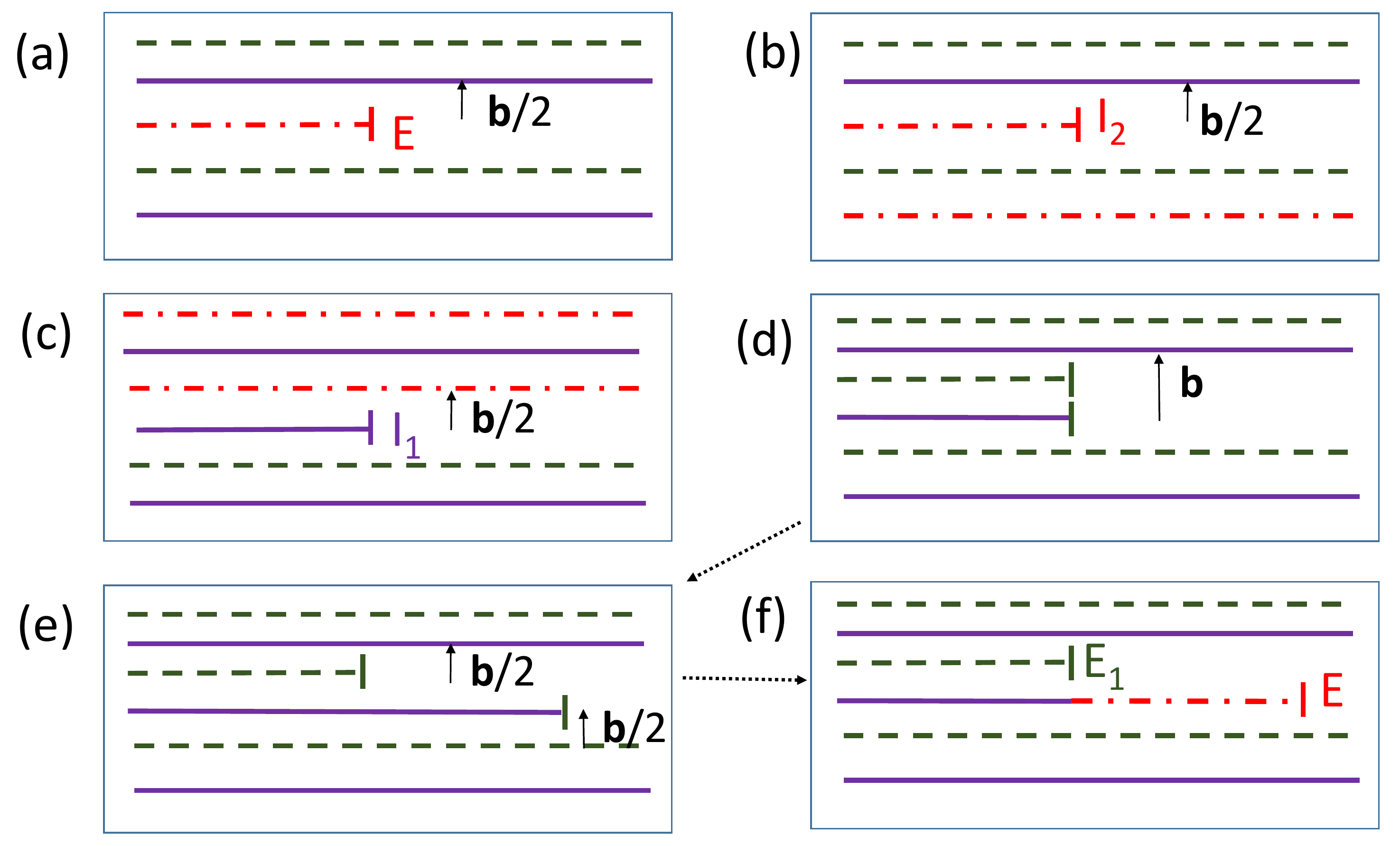}
\caption{ (Color online) Partial dislocations as edges of the faults including those arising from splitting 
of the full dislocation. The layers A,B,C are shown respectively by dashed, solid and dot-dashed lines. 
(a) The $E$ partial; 
(b) The $I_2$ partial; 
(c) The $I_1$ partial; 
(d) The full edge  dislocation; 
(e) splitting of the full edge dislocation; 
(f) The final configuration of the split full dislocation.   }
\label{fig3}
\end{figure}

The hcp crystal of \he4 can be viewed as the ABABAB... stack of identical triangular layers A and B shifted 
with respect to each other as explained in Fig.~\ref{fig1}.
At high pressure there is a transition to the fcc phase which corresponds to the ABCABCABC... stacking, see Fig.~\ref{fig1}.
The full edge and screw dislocations possessing superfluid cores as found in Refs.~\cite{sclimb,screw}, are shown schematically 
in Fig.~\ref{fig2}. The structure of the edge partials is shown schematically in Fig.~\ref{fig3}.

\section{{\it Ab initio} simulations}\label{sec2}

First principles studies of small \he4 samples we perform using the worm algorithm \cite{WA} by closely following 
the scheme described in previous works \cite{screw,sclimb}.
\begin{figure}[!htb]
\includegraphics[width=1.0 \columnwidth]{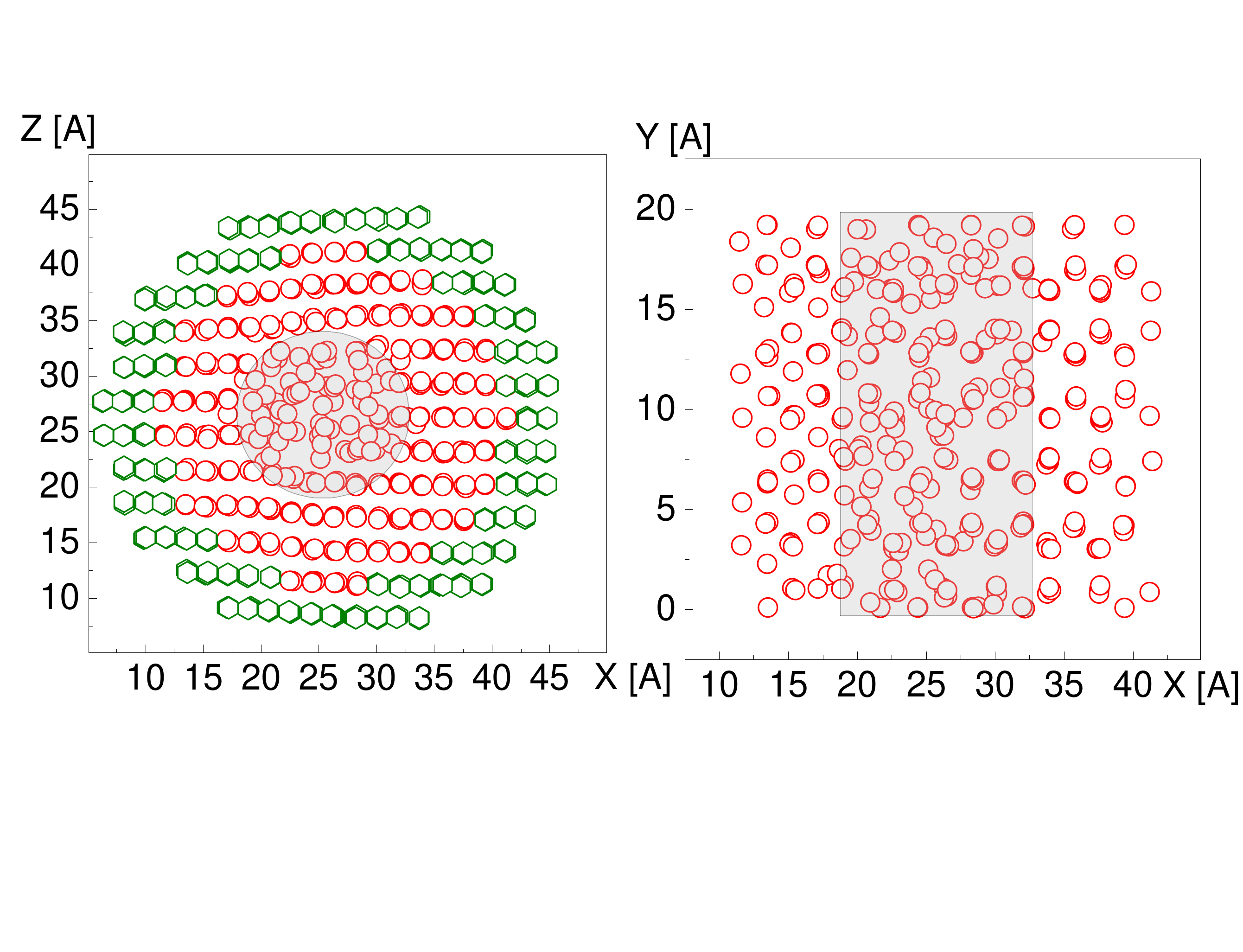}
\vskip -0.8cm
\caption{ (Color online) A snapshot of atomic positions in a typical sample containing the partial dislocation 
with its core along the Y-axis. Fully updatable (382) particles are shown by open (red) circles. 
The open (green) hexagons are frozen particles. The shaded areas indicate the region of the superfluid core 
(of the radius $\approx 7$\AA ) where atoms appear to be disordered by participating in exchange cycles.
Left panel: the columnar view along the core (Y-axis). Right panel: the columnar view along the hcp axis 
(frozen particles are not shown).    }
\label{fig4}
\end{figure}
\begin{figure}[!htb]
\includegraphics[width=0.95 \columnwidth]{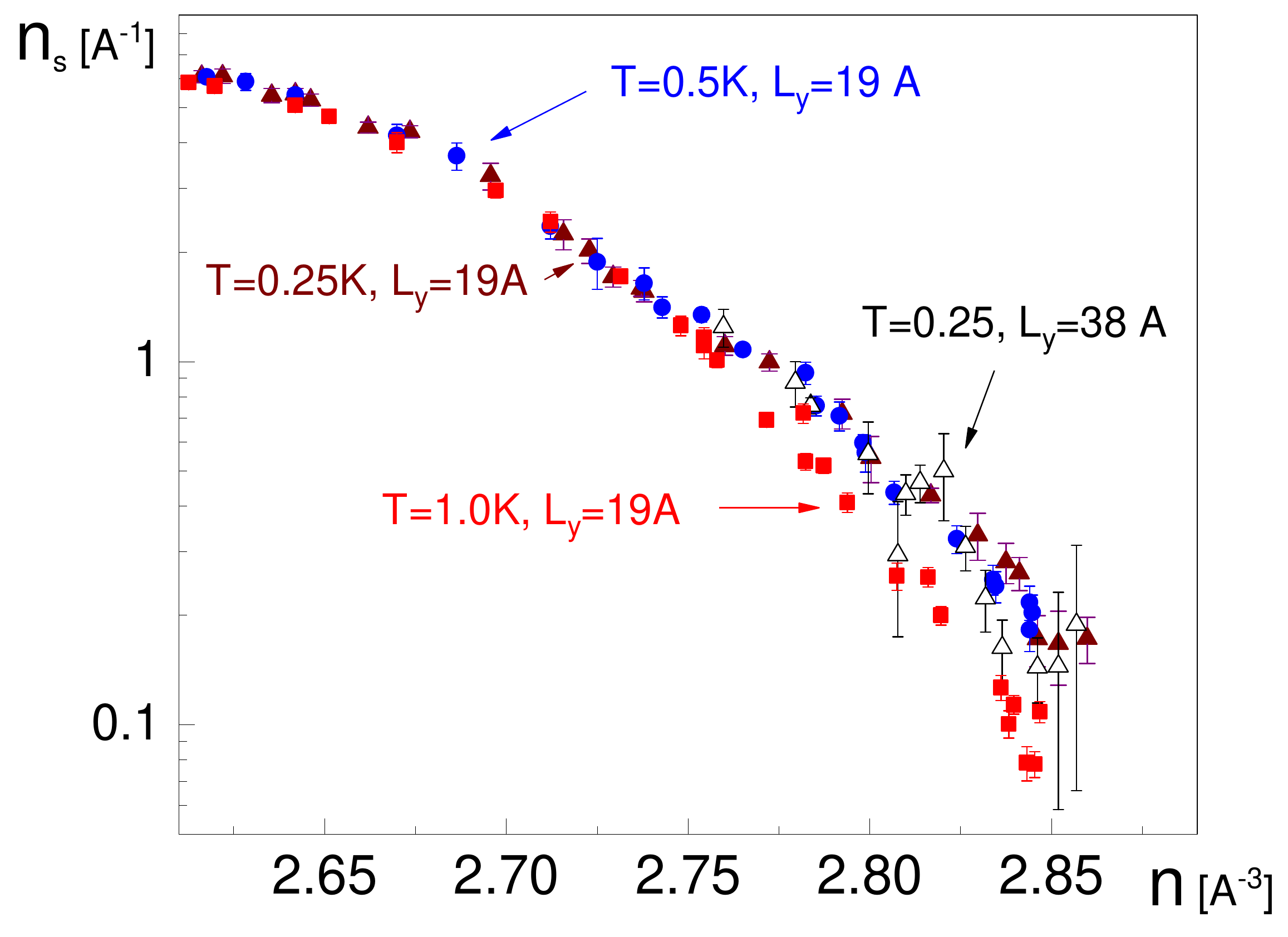}
\caption{ (Color online) Superfluid stiffness $n_s$, Eq.~(\ref{rho_kap}), as a function of $n$
for the $E_1$-partial at several temperatures and two core lengths: 
$T=0.25K, \, L_y=19\AA$ (filled wine triangles); 
$T=0.25K, \, L_y=38\AA$ (open triangles);  
$T=0.5K, \, L_y=19\AA$ (filled blue circles); 
$ T=1.0K, L_y=19 \AA$ (filled red squares).  }
\label{fig8}
\end{figure}
The initial configuration was prepared starting from atomic positions in the ideal hcp solid. 
In order to produce a partial edge dislocation the corresponding half of the atomic plane was removed and
the configuration was annealed by purely classical simulations in the canonical ensemble. 
Given periodic boundary conditions (PBC), the sample contained two partials with opposite orientations of the Burgers vectors. 
In the grand canonical worm algorithm simulations these partials would be eliminated by either completely removing 
the incomplete atomic layer or by extending it over the whole sample.
To prevent this from happening, all atoms outside the concentric cylinder of the radius $R_0=15\AA$ around the core located 
at the center of the sample have been frozen (after updating particle world lines in the canonical ensemble by treating them as distinguishable). Full quantum mechanical simulations with quantum exchanges and in the grand canonical ensemble 
were performed only for the particles inside the cylinder.  
A typical snapshot of the atomic configuration is shown in Fig.~\ref{fig4}.
The superfluid response is concentrated in the shaded areas surrounding the core.  
We did not detect any finite size effects in the density dependence of $n_s$ by simulating samples with 
the dislocation length $L=19\AA$ and $L=38\AA$.
It should be mentioned that the frozen particles at the periphery of the sample slightly shift 
the melting line to a lower density.

 Screw dislocation samples were prepared similarly---by adding 
the screw type displacement to the ideal hcp structure. 
In order to mitigate the effect of the non-PBC in the directions perpendicular  
to the dislocation, the wall of the frozen particles has been created as well, with its symmetry axis aligned
with the Z-axis. Accordingly, the PBC were preserved along this axis.

Two key parameters characterizing the hydrodynamic response are the superfluid stiffness $\Lambda$ and the 
compressibility $\kappa$ defined as \cite{Pollock_Ceperley}
\beq
\Lambda =  \langle W^2\rangle  LT, \qquad 
\kappa  =  \langle W^2_\tau \rangle/(LT) ,
\label{rho_kap}
\eeq
where $W$ and $W_\tau$ are the space and time winding numbers of the particle worldlines, respectively.  
The superfluid density and stiffness are related by $n_s= m\Lambda/\hbar^2$ with $m$ being the mass of the \he4 atom.
In one dimensional systems, an important quantity is the Luttinger parameter
\beq
K=\pi \sqrt{\Lambda \kappa }= \pi \sqrt{\langle W^2\rangle \langle W_\tau^2 \rangle}.
\label{K}
\eeq
In the Luttinger liquid theory, it determines the stability of the superfluid phase against 
disorder and commensurate potentials with the typical critical value $K_c =2$ for the latter below which
the insulating behavior emerges.

 At this juncture we note that, while the screw dislocation with the superfluid core \cite{screw} is well described within the Luttinger liquid paradigm with finite $K$ well above its critical value, the situation with the superclimbing edge dislocation \cite{sclimb} is radically different. Here $K \propto L $ diverges with dislocation length $L\to \infty$. 
This behavior is the indication that the Luttinger liquid description does not apply in general to the superclimbing dislocations, and observing finite $K$ in {\it ab initio} simulations is a consequence of the dislocation confinement within the wall of frozen particles in a finite-size sample. Conversely, if the special condition for suppressing the quantum roughness is realized as described in Refs.~\cite{Max}, the superclimb can vanish at $T\to 0$ and, accordingly, the Luttinger liquid behavior will be restored in the thermodynamic limit.

Simulation results for $n_s$ of the edge and the screw dislocations are presented in Figs.~3 and 4  of the main text and in Fig.~\ref{fig8} here.
It is worth noting that $n_s$ changes by more than one order of magnitude, while in the regime of strong superfluid (with $K>2$) the sample density changes by only a few percent. 
At the same time, no significant temperature (up to $T=1$K) or dislocation length dependencies were observed
in our samples, see Fig.~\ref{fig8}.
There is a difference between the protocols of how Fig.~3 in the main text and Fig.~\ref{fig8} shown here were obtained.
In the former, the edge dislocation was kept at the center of the sample prepared at some initial density by 
fine-tuning the chemical potential.   
In the latter, Fig.~\ref{fig8}, the edge dislocation in the same sample (for each $T$ and $L$) was pushed 
towards the perimeter formed by the frozen particles by increasing $\mu$. This resulted in varying the 
sample density without changing significantly the density inside the core.

As emphasized in the main text, $n_s$ of the screw dislocation is sensitive to the density of the crystalline environment rather than
to the density inside its core. As can be seen from Fig.~4 of the main text, 
changing the particle number inside the core does not affect $n_s$ of the screw dislocation.
At the same time, changing the overall sample density does suppress $n_s$ significantly (see Fig.~3 of the main text).

\section{Effective classical model for a superclimbing edge dislocation}\label{sec3}
Here we provide more details about  the simplified model (1) introduced in the main text.
Eq.~(3) of the main text can be written as
\beq
 V_0 + \varphi' = \frac{J}{n_0} \exp(g y'') \, .
\label{phi}
\eeq
Integrating both sides over $x$, given the condition $dJ/dx=0$, and taking the PBC into account, we find
\beq
J= \frac{n_0 V_0 L}{\int_0^L dx \exp(g y'')} \,.
\label{JV}
\eeq
Accordingly, the shape dependent energy functional (1) in the main text is given by
\beq
H=\frac{ n_0V^2_0 L^2 }{2\int_0^L dx \exp(g y'')} + \int_0^L dx  \frac{G}{2}(y')^2.
\label{E2}
\eeq
It has been discretized by introducing the lattice with the period $\Delta x$ which depends on temperature
as explained in the main text. Accordingly, the derivatives $y'$ and $y''$ were changed as
$y' \to \tilde{y}'=(y(i+1) - y(i))/\Delta x$ and $y'' \to \tilde{y}''=(y(i+2)+y(i)  - 2y(i+1))/(\Delta x)^2$, where $i=0,1,....\tilde{L}-1$ are sites of the 1D lattice. 
The classical Monte Carlo simulations were conducted with the weight $\exp( - H/T)$ by proposing random changes of $y(i)$ 
within the constraint $|\tilde{y}'(i)|<1$. The quantities $n_s$ and $J_c$ were determined as described in the main text.  
The velocity $V_0$ has been chosen well below the critical speed $V_c$ above which the structural instability develops.

\subsection{Structural instability and the critical current. }

Straightforward analysis of  Eq.~(\ref{E2}) reveals its structural instability at finite $V_0$.
Indeed, by expanding (\ref{E2}) in small $y'\to 0$ and $y''\to 0$ one finds
Eq.(\ref{E2}) as
\beq
H\approx \int_0^L dx \left[ - \frac{g^2 n_0 V_0^2}{4} (y'')^2 +   \frac{G}{2}(y')^2\right],
\label{E3}
\eeq
where the linear term $\sim \int dx y'' =0$ due to the periodic boundary condition.
Since the first term here is negative, all Fourier harmonics of $y$ with momenta 
\beq
q > q_c = \sqrt{\frac{2G}{n_0 V_0^2 g^2}}
\label{inst}
\eeq 
become unstable at finite $V_0$.
However, the effective long-wave model is only valid at length scales much larger than the core diameter $D_0$, 
or, equivalently, at wave vectors $q \ll 1/D_0$. Thus, the instability emerges only at large enough $V_0$  
\beq
V_0 > V_c \gg \sqrt{\frac{2G}{n_0}}\frac{D_0}{g}.
\label{Vc}
\eeq
This instability, which can be traced back to the Landau criterion, reduces the critical superflow velocity roughly by a factor 
$\sim  g^{-1} \ll 1$ as compared to the speed of sound in solid \he4.

In the discrete classical version of the model with the lattice period $\Delta x \gg D_0$, the discrete momenta
are given as $q= 2\pi m/(\Delta x \tilde{L})$, where $L = \tilde{N} \Delta x$ and  $m=0, 1, 2, ... , \tilde{L} -1$. 
Accordingly, the discrete Fourier images of $y''$ and $y'$ become $- 4 \sin^2(\pi m/ \tilde{L}) y_m/(\Delta x)^2$ and
$2i\sin(\pi m/\tilde{L}) y_m/\Delta x$, respectively, with $y_m$ being the discrete Fourier image of $y(i)$.  The critical  
speed, then, becomes
\beq
V_c = \sqrt{\frac{G}{2n_0}}\frac{\Delta x}{g}.
\label{Vc2}
\eeq
It corresponds to the unstable harmonic with $m=\tilde{L}/2$, that is, $q_c=\pi/2$.

In this work, we assume that the critical current is limited by quantum phase slips occurring 
at local velocities well below $V_c$. The phenomenological condition is expressed by Eq.~(4)
in the main text where $p>1$. This choice is motivated by the experimental data of Ref.~\cite{Moses}
showing that the exponential dependence of $F$ on pressure is much steeper than what we 
observe for $n_s$ in {\it ab initio} simulations presented in Fig.~3 (by a factor 5 -- 10) 
and in Figs.~4 of the main text and Fig.~\ref{fig8} here (by a factor of 2 -- 5).

\subsection{The dependence of $n_s$  on the dislocation line curvature}
We start by relating the line curvature to the local change in the crystal density, $\delta n(x)$,  
in the vicinity of the core \cite{Lothe}. 
In the discretized version of the model, this relation is given by 
\beq
\delta n(x)= \tilde{b} y''(x), \qquad \tilde{b} = \frac{(1-2\nu)b}{2\pi(1-\nu)} \ln\left[\frac{\Delta x}{D_0}\right].
\label{Phi}
\eeq
where $\nu$ stands for the Poisson coefficient, $b$ is the Burgers vector, and  
$D_0$ is the short distance cutoff taken to be the dislocation core width $D_0$ (of about 10$\AA$). 
The overall sign depends on the Burgers vector direction $b$ and the direction of the displacement $y$---that is, 
whether atoms are added to or subtracted from the incomplete atomic plane.
{\it Ab initio} simulations established that $n_s$ dependence on the density variation is 
exponential, see Eq.~(2) in the main text. With the help of Eq.~(\ref{Phi}) 
we relate the exponential slope $\tilde{g}$ of the $n_s$ vs $\delta n$ dependence to the coefficient $g$
(introduced in Eq.(2) of the main text) 
as
\beq
g= \frac{(1-2\nu)b}{2\pi(1-\nu)} \ln\left[\frac{\Delta x}{D_0}\right]\tilde{g}.
\label{g}
\eeq
In our system of units, $\hbar=1, m=1, b=1, k_B=1$, 
the values of $\tilde{g}$ in Figs.~3 and 4 of the main text are roughly $\tilde{g} \sim 40-50$
and $\tilde{g} \sim 70-100$, respectively ($\tilde{g} \sim 40-50$ also in Fig.~\ref{fig8} above). 
Equation (\ref{g}) then predicts that $g\approx 6-10$. In the simulations we have used a fixed value of 
$g=40/2\pi \approx 6.37$ for the entire temperature range (that is, we dropped the factor $\frac{1-2\nu}{1-\nu}\ln(\Delta x/D_0)$ in Eq.(\ref{Phi})).

As for other parameters, the line tension $G$ is given by the $C_{44}$ modulus of the solid \he4
as $G\approx C_{44}b^2/4\pi$ and in our units corresponds to temperature $\approx 2$K \cite{elast}. 
The value of $n_0$ was chosen from the middle of the data set shown in Fig.~\ref{fig8}, $n_0 \approx 1-2 \AA^{-1}$,
or $5-10$K in the chosen units. The simulation data presented in the main  text correspond to $n_0=10$K.

\subsection{Superfluid density and critical current.}
The dependence $\ln n_s \propto - T^\alpha$ with $\alpha=5/2$ shown in Fig.~5 of the main text can be understood as follows.
In the steady flow regime at small $V_0$, the kinetic energy is small compared to the elastic  energy 
$\sim G  \int dx \, (y')^2$. Accordingly, the averaging of $\int dx \, n^{-1}_s(x)$ over fluctuations of $y$ results in 
$n_s \approx n_0  \exp[ -g^2 \langle (y'')^2\rangle/2] $. The value of $\alpha=5/2$ follows directly from 
$ \langle (y'')^2\rangle \sim T^{5/2}$, and the typical temperature scale is given by 
$T_\alpha \sim T_0^{1/\alpha}$ where $T_0 =G^{7/4} n_0^{3/4}g^{-2}\approx 0.47$K for the chosen parameters. 

The weakest region (WR) along the dislocation corresponds to a spike in $y''>0$  
such that the local $n_s$ (see Eq.(2) in the main text) is strongly suppressed. 
This can be realized despite the conditions $|y'|<1$, $|y''|<1/D$ as long as $g$ is large enough. 
Within the  model (\ref{E2}), a simplified picture of the WR is based on the 
shape fluctuation such that $y(i)=0$ everywhere except for  some site $i=x_0$, so that $y(x_0) = -Y <0$. Accordingly,
$\tilde{y}'(x_0)= Y/\Delta x,\,\, \tilde{y}'(x_0-1)= - Y/\Delta x$, $\tilde{y}''(x_0-2) =- Y/(\Delta x)^2,\,\, \tilde{y}''(x_0-1) = 2Y/(\Delta x)^2$, and $\tilde{y}''(x_0)= - Y/(\Delta x)^2$ 
(all other values are zero). Substituting these values into Eq.~(\ref{E2}) we observe that the 
kinetic energy term is exponentially small, $\propto \exp\left[ - 2gY/(\Delta x)^2 \right] \ll 1$ at $i=x_0-1$,
while the elastic one, $\approx G Y^2/\Delta x$, dominates in the configuration energy.

The probability for having large $Y$ anywhere on the dislocation line can be estimated as 
$\approx \tilde{L} \exp \left[ - G  Y^2/(\Delta x T) \right]$. 
This probability is of order one for
\beq
Y \approx \Delta x \sqrt{ \ln(\tilde{L})\, T/G\Delta x } .
\label{YT}
\eeq
The condition $|y'|<1$ [used in Eqs.~(1)--(3) of the main text] implies that $Y$ cannot exceed $\Delta x$.
 
The statistical average $\langle ... \rangle_{\rm WR}$ over WR in Eq.~(4) of the main text 
has been conducted for $p=2$ according to the following
protocol: A configuration $y(x)$ contributes to the average if the following condition
\beq
\tilde{y}''(i) > \gamma \frac{Y}{(\Delta x)^2}
\label{YxxYT}
\eeq
is satisfied. The coefficient $\gamma =0.3$ was chosen such that the inequality (\ref{YxxYT}) was satisfied 
at about 10-20\% of the sites.
We verified that increasing $\gamma$ by a factor of $3$ in order to decrease the number of WR sites to just a few instances
changes the characteristic temperature scale $T_\alpha$ by a factor of $2$, but does not change the 
power $\alpha\approx 5/4$ in the temperature dependence  $\ln J_c = - (T/T_\alpha)^\alpha$.

\subsection{The bulk phonon mechanisms as a source of the fluctuations.}
As mentioned in the main text, at $T>0.5-0.6$K bulk phonons start making a significant contribution to 
fluctuations of $\delta n$ along the dislocation line, and, thus, suppress $n_s$. 
This contribution is estimated as $\langle (\delta n)^2\rangle \approx (T/T_D)^4$ where
$T_D$ is of the order of the Debye temperature $T_D\sim 30$K. This effect alone would predict 
$\ln n_s \sim - \tilde{g}^2 (T/T_D)^4$ for the equilibrium superfluid density and 
$\ln J_c \sim - p\tilde{g} (T/T_D)^2\sqrt{\ln \tilde{L}}$ for the critical current. 
It appears that a crossover from $\alpha =5/4$ to $\alpha=2$ can explain the deviation of the experimental data points 
from the model fits at high temperature, see Fig. 1 in the main text.   
It is also worth mentioning that the bulk phonons mechanism will affect both the edge and screw dislocations.


\begin{thebibliography}{99}

\bibitem{Hallock} M. W. Ray and R. B. Hallock, Phys. Rev. Lett. {\bf 100}, 235301
(2008); Phys. Rev. {\bf B 79}, 224302 (2009); M. W. Ray and R. B. Hallock, Phys. Rev. {\bf B 81}, 214523 (2010).

\bibitem{Hallock2012} Ye. Vekhov and R. B. Hallock,
Phys. Rev. Lett. {\bf 109}, 045303 (2012).

\bibitem{Hallock2019} R.B. Hallock, J. Low Temp. Phys. {\bf 197}, 167 (2019)%

\bibitem{Beamish} Z. G. Cheng, J. Beamish, A. D. Fefferman, F. Souris, S. Balibar, and V. Dauvois,
Phys. Rev. Lett. {\bf 114}, 165301 (2015);  Z. G. Cheng and J. Beamish, Phys. Rev. Lett. {\bf 117}, 025301 (2016).

\bibitem{Moses} J. Shin, D. Y. Kim, A. Haziot, and M. H. W. Chan, Phys. Rev. Lett. {\bf 118}, 235301 (2017).

\bibitem{Moses2019} J. Shin and M. H. W. Chan, Phys. Rev. {\bf B 99}, 140502(R) (2019).

\bibitem{Moses2020} J. Shin and M. H. W. Chan, Phys. Rev. B {\bf 101}, 014507 (2020).

\bibitem{Moses2021} M.H.W. Chan, J. Low Temp. Phys. {\bf 205}, 235 (2021).

\bibitem{Beamish_Balibar} A. Haziot, X. Rojas, A. D. Fefferman, J. R. Beamish, and S. Balibar,
Phys. Rev. Lett. {\bf 110}, 035301 (2013).

\bibitem{screw}  M. Boninsegni, A. B. Kuklov, L. Pollet, N. V. Prokof'ev, B. V. Svistunov, and M. Troyer, Phys. Rev.Lett. {\bf 99}, 035301 (2007).


\bibitem{sclimb} S. G. S\"oyler, A. B. Kuklov, L. Pollet, N. V. Prokof'ev, and B. V. Svistunov, Phys. Rev. Lett. {\bf 103}, 175301(2009).

\bibitem{shevchenko} S. I. Shevchenko, Fiz. Nizk. Temp. {\bf 13}, 115 (1987) [Sov. J. Low Temp. Phys. {\bf 13}, 61 (1987)].


\bibitem{Lothe}  P. M. Anderson,  J. P.  Hirth,  and J. Lothe, {\it Theory of Dislocations}, Cambridge University Press, 2017.

\bibitem{Hull} D. Hull, D. J. Bacon, {\it Introduction to Dislocations}, Elsevier, Amsterdam-Tokyo, 2011.

\bibitem{SM} See the Supplemental Material for details of the dislocation structure, sample preparation, and simulations.

\bibitem{trends}  A.B. Kuklov, N.V. Prokof'ev, and B.V. Svistunov,  {\it How Solid is Supersolid?},  Physics {\bf 4}, 109 (2011).



\bibitem{WL00}  E. Altman, Y. Kafri, A. Polkovnikov, and G. Refael, Phys.
Rev. Lett.  {\bf 93}, 150402 (2004); ibid Phys. Rev. B {\bf 81}, 174528 (2010).

\bibitem{WL0}  L. Pollet, N.V. Prokof'ev, and B.V. Svistunov,
Phys. Rev. {\bf B 89}, 054204 (2014).

\bibitem{WL} Z. Yao, L. Pollet, N. V. Prokof'ev, B. V. Svistunov, New J. Phys. {\bf 18}, 045018 (2016). 



\bibitem{Langer} J. S. Langer and V. Ambegaokar, Phys. Rev.  {\bf 164}, 498 (1967).

\bibitem{WA} M. Boninsegni, N. V. Prokof'ev, B. V. Svistunov, Phys. Rev. Lett. {\bf 96}, 070601 (2006); Phys. Rev. {\bf E 74}, 036701 (2006).


\bibitem{Pollock_Ceperley} E. L. Pollock and D. M. Ceperley, Phys. Rev. B {\bf 36}, 8343 (1987).

\bibitem{Max} 
M. Yarmolinsky and A. B. Kuklov, Phys. Rev. {\bf B 96}, 024505 (2017);
Longxiang Liu and A. B. Kuklov, Phys. Rev. {\bf B 97}, 104510 (2018). 

\bibitem{elast} D. S. Greywall, Phys. Rev. {\bf B 16}, 5127 (1977). 
\end{thebibliography}
\end{document}